\documentclass[twocolumn]{aastex63}

\received{April 4th, 2021}
\accepted{September 7th, 2021}
\submitjournal{ApJS}

\shorttitle{A Grid of Synthetic Spectra for Subdwarfs}
\shortauthors{Pacheco et al.}
\graphicspath{{./}{figures/}}

\begin{document}

\title{A Grid of Synthetic Spectra for Subdwarfs: Non-LTE line-blanketed atmosphere models}

\correspondingauthor{Thayse A. Pacheco}
\email{thayse.pacheco@usp.br}

\author[0000-0002-8139-7278]{Thayse A. Pacheco}
\affiliation{Universidade de São Paulo, Instituto de Astronomia, Geofísica e Ciências Atmosféricas, Rua do Matão, 1226 São Paulo, SP 05508-900, Brazil}

\author[0000-0002-6040-0458]{Marcos P. Diaz}
\affiliation{Universidade de São Paulo, Instituto de Astronomia, Geofísica e Ciências Atmosféricas, Rua do Matão, 1226 São Paulo, SP 05508-900, Brazil}

\author[0000-0003-2499-9325]{Ronaldo S. Levenhagen}
\affiliation{Universidade Federal de São Paulo, Departamento de Física, Rua Prof. Artur Riedel, 275, CEP 09972-270, Diadema, SP, Brazil}

\author[0000-0003-1846-4826]{Paula R. T. Coelho}
\affiliation{Universidade de São Paulo, Instituto de Astronomia, Geofísica e Ciências Atmosféricas, Rua do Matão, 1226 São Paulo, SP 05508-900, Brazil}



\begin{abstract}
A new grid of detailed atmosphere model spectra for hot and moderately cool subdwarf stars is presented. High-resolution spectra and synthetic photometry are calculated in the range from 1000 \texttt{\AA} to 10,000 \texttt{\AA} using Non-LTE fully line-blanketed atmosphere structures. Our grid covers eight temperatures within 10,000 $\le$ $T\textsubscript{eff}$ [K] $\le$ 65,000, three surface gravities in the range 4.5 $\le \log{g}$ [cgs] $\le$ 6.5, two helium abundances matching two extreme helium-rich and helium-poor scenarios, and two limiting metallicity boundaries regarding both solar ([Fe/H] = 0) and Galactic halo ([Fe/H] = -1.5 and [$\alpha$/Fe] = +0.4). Besides its application in the determination of fundamental parameters of subdwarfs in isolation and in binaries, the resulting database is also of interest for population synthesis procedures in a wide variety of stellar systems.

\end{abstract}



\section{Introduction} \label{sec:intro}

The evolutionary scenario of subdwarfs is still emerging. Hot subdwarf stars evolve from low- to intermediate-mass progenitors and reach far beyond the main sequence, at the blue end of the horizontal branch (HB) or mixed with post-HB stars \citep{Heber2009}. Even  though they lie mostly between the main sequence and the white dwarf (WD) cooling sequence in the Hertzsprung-Russell diagram, they are  contaminated with WDs.

The most accepted formation scenario is of an extreme HB star that lost its envelope after the He-burning phase, evolving directly to the WD-cooling sequence by avoiding the asymptotic giant branch (AGB) \citep{Wade2004,Heber2009, Fontaine2012}. These objects experienced mass transfer followed by common envelope ejection in a binary system leaving the stellar core exposed \citep{Green2008, Geier2010}. More recent investigations support this scenario, such as the observational evidence of hot O- and B-type subdwarf formation (sdO and sdB, respectively) from the product of binary interactions \citep{pelisoli2020}. Also, the expected evolutionary path suggests that sdOs evolve from sdBs \citep{Heber2009}. An alternative evolutionary scenario suggests that the merger of two low-mass He-core WDs might form isolated H-rich B-type hot subdwarfs \citep{Hall2016,Schwab2018}.

A small fraction of sdBs in close binaries \citep{Wade2004} fall out of the red giant branch before He ignition, perhaps because they are low-mass stars that do not support the burning of He in their cores \citep{Heber2009}. 

The hot subdwarfs can be separated into three main groups according to \cite{saffer00}. i) Those nonbinaries that have no detectable spectral lines from a cool companion, and show only small or insignificant velocity variations ($35\%$). ii) Those single-lined spectroscopic binaries which have significant or large velocity variations and probable orbital periods on the order of a day ($45\%$). iii) Those showing additional spectral lines from a cool FGK main-sequence or subgiant companion that have slowly varying or nearly constant velocities, indicating periods of many months to years. 

There is observational evidence indicating that subdwarfs are the most likely candidates to account for the ultraviolet (UV) upturn observed on spectroscopic and photometric analyses of globular clusters and elliptical galaxies \citep{Yi1999, Busso2005, Green2008}.

Analysis from asteroseismology \citep{Charpinet2005}, spectroscopy \citep{Wade2004, Dorsch2017, Dorsch2019}, and photometry \citep{Johnson2013} of field subdwarf stars found nonsolar He abundances. He-rich and He-poor sequences, as a function of effective temperature, were proposed by \citet{Edelmann2003} and updated by \citet{nemeth2012} and \citet{Lei2019}. Those sequences also support the evidence that He-enriched sdOs are more common than the He-deficient ones \citep{Heber2009}. 

It is ideal to have homogeneous and widely available model spectra for these elusive systems, in order to improve our knowledge of their fundamental physical parameters and evolution. Such models should be computed as fully as possible regarding the radiative transfer,
considering non-local thermodynamic equilibrium (NLTE), fully blanketed atmosphere structures \citep{metalNLTE1995, Lanz1997}. 

Considerable improvements in NLTE atmosphere models have been achieved over the last five decades. The first NLTE spectra of hot atmospheres in the early 1970s \citep{mihalas74} were built based on pure H atmosphere structures, and the metal lines were considered only in the spectral synthesis. Analysis by \citet{metalNLTE1995} took H, C, and Fe into account for the opacity of hot and high-gravity NLTE atmosphere models. They found critical differences in effective temperature determinations and line profiles arising from the inclusion of explicit atomic species (i.e. with the kinetic equilibrium equation solved) in the atmospheric structure. The subdwarfs' atmospheres have more evident NLTE effects due to their lower surface gravities (when compared to WDs; \citealt{Lanz1997}). They built spectra considering full line blanketing with H/He, CNO, Si, Fe, and Ni as implicit ions and concluded that this approach significantly improved the line profile analysis. \citet{nlte1998} showed that NLTE metal-line-blanketing effects produce a conspicuous difference (typically 20$\%$) in the line profiles seen in the UV spectra of hot WDs. By considering explicit metal ions such as H/He, CNO, Si, and Fe in the atmosphere structure a better emergent flux match was achieved. 

Overall, the subdwarfs' spectra are built for specific targets and studies. We have hundreds of hot subdwarf stars cataloged \citep{Geier2014, Heber2017, Geier2019, Geier2020}, with their spectral analysis focused on classification and/or kinematic properties \citep{Luo2019}. A lot of work was done to analyse the He-abundance \citep{Fontaine2019} sequences \citep{Lei2019}. 
The sequences also show differences between field subdwarfs and extreme HB subdwarfs in globular clusters \citep{Lei2020}.

NLTE structures and subdwarf spectra were calculated by \citet{Nemeth2014}, covering specific temperature and gravity ranges. The opacities of H, He, and a few metal ions  were included to compute the atmosphere structure in these models. However, only a small and nonhomogeneous collection of detailed subdwarf model spectra is currently available. High-temperature spectral grids may be also calculated by other codes e.g. the Tübingen NLTE Model Atmosphere package \citep{Rauch2018}\footnote{\url{http://dc.zah.uni-heidelberg.de/theossa/q/web/form}}.

In this work we present an extensive high-spectral-resolution grid of NLTE synthetic spectra for subdwarfs in the optical and UV. The models are fully line blanketed with H, He, C, N, O, Ne, Mg, Al, Si, S, and Fe as opacity sources (see table \ref{ions}). We also consider nonsolar chemical abundances to better sample the subdwarf parameter space. Low-temperature convective atmospheres were also included in the square grid. The complete grid is available on our SpecModel website \footnote{Spectral Models of Stars and Stellar Populations, \url{http://specmodels.iag.usp.br/}.} as well as on the SVO's Theoretical Spectra Web Server \footnote{Spanish Virtual Observatory, \url{http://svo2.cab.inta-csic.es/theory//newov2/}.} and the Vizier \footnote{\url{https://vizier.u-strasbg.fr/viz-bin/VizieR}} databases.

Section \ref{sec:atm} describes the atmosphere structure models, in particular the differences between the LTE and NLTE assumptions made. Section \ref{sec:spec} describes the synthetic spectra, comparing the results with observational data. Section \ref{sec:mag} describes the construction of the synthetic magnitudes and their application in color-color diagrams. Section \ref{sec:conc} presents the concluding remarks.

\section{Atmosphere Structure} \label{sec:atm}

The atmosphere structure models were computed by \texttt{TLUSTY} code v205 and v208 (for convective 10,000 K models; \citealt{hubeny1988, tlusty2011}). It calculates a self-consistent solution of the equations describing the radiative (or radiative plus convective) transfer and physical state of the atmosphere. The geometry of the model is  plane-parallel with homogeneous chemical abundances \citep{CLALI1995}. 

The most popular subdwarf models available have been built in LTE. In LTE all energy partitioning such as atomic, ionic, and molecular level populations is given by Saha-Boltzmann equations, being defined by the local temperature. The LTE conditions may differ from those derived in actual statistical equilibrium, therefore level populations in NLTE were calculated. Departures from LTE may be specially significant for high-temperature subdwarf atmospheres \citep{CLALI1995, metalNLTE1995, Lanz1997} as also found in DA-type WD atmospheres \citep{Levenhagen2017}. 

\begin{table*}[hb!]
\centering
\caption{Atomic data of the explicit species included in the NLTE atmosphere models}
\begin{tabular}{lccc|cccccccc}
\hline \hline
\multicolumn{4}{c}{\textsc{Atomic data}} & \multicolumn{8}{c}{\textsc{Explicit Ions included in NLTE}}\\
\hline \hline
Ion & Super(level) & Lines & Reference & 10 kK & 15 kK & 20 kK & 25 kK & 30 kK & 35 kK & 45 kK & 65 kK\\
\hline

 H$^{-}$ & 1 & 1 & 	1	& \checkmark& \checkmark& \checkmark& & & & &\\
\hline
 H & 9 & 172 & 	1			& \checkmark& \checkmark& \checkmark&  \checkmark& \checkmark&  \checkmark& \checkmark& \checkmark\\
\hline
 He  & 24 & 784 & 2 & \checkmark& \checkmark& \checkmark&  \checkmark&  \checkmark& \checkmark& \checkmark& \checkmark\\ 
 He \textsc{ii} & 20 & 190 & 1		& \checkmark& \checkmark& \checkmark&   \checkmark& \checkmark& \checkmark& \checkmark& \checkmark\\
\hline
 C & 40 & 3,201 & 3			& \checkmark& \checkmark& \checkmark&  \checkmark& & & &\\ 
 C \textsc{ii} & 22 & 238 & 4		& \checkmark& \checkmark&  \checkmark& \checkmark&  \checkmark& \checkmark& \checkmark& \checkmark\\
 C \textsc{iii} & 46 & 738 & 5		& \checkmark& \checkmark&  \checkmark& \checkmark&  \checkmark& \checkmark& \checkmark& \checkmark\\
 C \textsc{iv} & 25 & 330 & 6	& & & & \checkmark& \checkmark& \checkmark&  \checkmark& \checkmark\\
\hline
 N & 34 & 785 & 1			& \checkmark& \checkmark& \checkmark& \checkmark&   & & &\\ 
 N \textsc{ii} & 42 & 3,396 & 3		& \checkmark& \checkmark&  \checkmark&  \checkmark& \checkmark& \checkmark& \checkmark& \checkmark\\
 N \textsc{iii}& 32 & 549 & 4		& \checkmark& \checkmark& \checkmark&  \checkmark&  \checkmark& \checkmark& \checkmark& \checkmark\\
 N \textsc{iv} & 48 & 1,093 & 5	& & & &\checkmark& \checkmark&  \checkmark& \checkmark& \checkmark\\
 N \textsc{v} & 16 & 330 & 6	& & & &\checkmark& \checkmark&  \checkmark& \checkmark& \checkmark\\
\hline
 O & 33 & 418 & 	1		& \checkmark& \checkmark& \checkmark&  \checkmark&  & & &\\  
 O \textsc{ii} & 48 & 3,484 & 1		& \checkmark& \checkmark&   \checkmark& \checkmark& \checkmark& \checkmark& \checkmark& \checkmark\\
 O \textsc{iii}& 41 & 3,855 & 3		&  & \checkmark& \checkmark&  \checkmark&  \checkmark& \checkmark& \checkmark& \checkmark\\
 O \textsc{iv} & 39 & 922 & 4	& & & & \checkmark& \checkmark&  \checkmark&  \checkmark& \checkmark\\
 O \textsc{v} & 6 & 4 & 4 	& & & & \checkmark& \checkmark&  \checkmark&  \checkmark& \checkmark\\
\hline
 Ne & 35 & 2,715 & 7		& \checkmark& \checkmark& \checkmark&  \checkmark&  & & &\\ 
 Ne \textsc{ii} & 32 & 2,301 & 1		&  \checkmark& \checkmark&  \checkmark&  \checkmark& \checkmark& \checkmark& \checkmark& \checkmark\\
 Ne \textsc{iii}&  34 & 1,354 & 1	& & & &  \checkmark& \checkmark& \checkmark& \checkmark& \checkmark\\
 Ne \textsc{iv} &  12 & 38 & 1 & & & & \checkmark& \checkmark& \checkmark&  \checkmark& \checkmark\\
\hline
 Mg \textsc{ii} &  25 & 306 & 1		& \checkmark&  \checkmark&  \checkmark& \checkmark& \checkmark& \checkmark& \checkmark& \checkmark\\
\hline
  Al \textsc{ii} & 29 & 536 & 8		& \checkmark&  \checkmark& \checkmark&  \checkmark& \checkmark& \checkmark& \checkmark& \checkmark\\
  Al \textsc{iii} & 23 & 306 & 1		& \checkmark&  \checkmark& \checkmark&  \checkmark& \checkmark& \checkmark& \checkmark& \checkmark\\
\hline
  Si \textsc{ii} & 40 & 392 & 9 & \checkmark&  \checkmark& \checkmark&  \checkmark& & & &\\ 
  Si \textsc{iii}& 30 & 747 & 8		& \checkmark&  \checkmark& \checkmark&  \checkmark& \checkmark& \checkmark& \checkmark& \checkmark\\
  Si \textsc{iv} & 23 & 306 & 1		&  & \checkmark&  \checkmark& \checkmark&  \checkmark& \checkmark& \checkmark& \checkmark\\
\hline
 S \textsc{ii} & 33 & 4,166 & 1 & \checkmark&  \checkmark& \checkmark&  \checkmark& & & &\\  
  S \textsc{iii}& 41 & 3,452 & 10		& \checkmark&  \checkmark&  \checkmark& \checkmark& \checkmark& \checkmark& \checkmark& \checkmark\\
  S \textsc{iv} & 38 & 909 & 9		&  & \checkmark&  \checkmark&  \checkmark& \checkmark& \checkmark& \checkmark& \checkmark\\
  S \textsc{v}  & 25  & 1,171 & 8  & & & & \checkmark& \checkmark&  \checkmark& \checkmark& \checkmark\\
  S \textsc{vi} & 16  & 398 & 1 & & & & & & &\checkmark&  \checkmark\\
\hline
  Fe \textsc{ii} & 36 & 1,264,969 & 12, 13 & \checkmark&  \checkmark&  \checkmark& \checkmark& & & &\\ 
  Fe \textsc{iii}& 50 & 1,604,934 & 11, 13		& \checkmark&   \checkmark& \checkmark& \checkmark& \checkmark& \checkmark& \checkmark& \checkmark\\
  Fe \textsc{iv} & 43 & 1,776,984 & 11, 14	&  & \checkmark&  \checkmark&  \checkmark& \checkmark& \checkmark& \checkmark& \checkmark\\
  Fe \textsc{v}  & 42 & 1,008,835 & 11, 15 & & & & \checkmark&  \checkmark& \checkmark& \checkmark& \checkmark\\
  Fe \textsc{vi} & 32 & 40,298 & 11, 15 & & & & & \checkmark & \checkmark&  \checkmark& \checkmark\\
 \hline \hline 
\end{tabular}
\label{ions} \\
References:
(1) \citealt{Lanz2003, Lanz2007};
(2) \url{http://physics.nist.gov/ PhysRefData /ASD/index.html}; 
(3) \citealt{LuoPradhan1989}; 
(4) \citealt{Fernley1999}; 
(5) \citealt{Tully1990}; 
(6) \citealt{Peach1988}; 
(7) \citealt{Hibbert1994};
(8) \citealt{Butler1993}; 
(9) \citealt{Mendoza1995};
(10) \citealt{Nahar1993}; 
(11) \citealt{Kurucz1994};
(12) \citealt{Nahar1997}; 
(13) \citealt{Nahar1996}; 
(14) \citealt{Bautista1997}; 
(15) \citealt{Bautista1996}. 
\end{table*}

The starting point was to compute a LTE gray structure model, which is then used as an input to build the NLTE line-blanketed models \citep{hubeny1988,nlte1994}. We considered H, He, C, N, O, Ne, Mg, Al, Si, S, and Fe as explicit ions (see table \ref{ions}) contributing to the opacity. The set of ionization states for the selected atomic species were chosen on the basis of ionization energies, model-effective-temperatures and candidate-permitted transitions from \cite{willians95}. The atomic data of each ion included as an explicit species in the NLTE atmosphere models are summarized in table \ref{ions} (see details in \citealt{Lanz2003}, \citealt{Lanz2007}). 

As described by \cite{Lanz2007}, the LTE fluxes are about 10$\%$ higher than the NLTE predictions, and this is most noticeable in the near-UV. \cite{Lanz2007} also mentioned that the lower NLTE fluxes result from the overpopulation of the H \textsc{i} (\textit{n} = 2) level at the depth of formation of the continuum flux, hence implying a larger Balmer continuum opacity.

The adopted abundances are expressed as solar ([Fe/H] = 0.0; \citealt{asplund09}) or Galactic halo ([Fe/H] = -1.5 and [$\alpha$/Fe] = +0.4) as well as He-rich \citep{Edelmann2003} and He-poor abundances \citep{nemeth2012,Lei2019}.
Our choice of using solar and halo metallicity is not intended to describe subdwarf typical abundances. Instead, these solar and halo points just set a broad \textit{Z} range, probing its effect on the continuum and line profiles. It is well established that atmospheric subdwarf metal abundances have little or virtually no memory of their parent main sequence abundances \citep[e.g.][]{moehler2001, nemeth2012, byrne2018}. Metals at the subdwarf surface vary by large amounts depending on previous binary evolution and diffusion processes, leading to diverse abundance patterns (e.g. \citealt{OTooleandHeber2006}) with loose observational constraints. He-rich and He-poor abundance variations were chosen based on linear fits as a function of temperature based on \cite{Lei2019} (see table \ref{n_He}). Note that \texttt{TLUSTY} \citep{hubeny1988, tlusty2011} uses the solar chemical abundances as defined by \cite{grevesse} as default values, and we rearranged all the chemical species to the solar abundance as defined by \cite{asplund09}. The atmosphere structures (available upon request to the authors) may be interpolated in metallicity and specific modified metal abundances used in custom spectral synthesis.

\begin{table}[htb]
\centering
\caption{Numerical He Abundances}
\begin{tabular}{ccc}
\hline \hline
$T\textsubscript{eff}$ [K]  & log($n$\textsubscript{He}/$n$\textsubscript{H}) \textsubscript{poor}  & log($n$\textsubscript{He}/$n$\textsubscript{H}) \textsubscript{rich}       \\
\hline 
10,000 & -4.98  &  -3.81    \\
15,000 & -4.35  & -3.13    \\
20,000 & -3.72  &  -2.46   \\
25,000 & -3.09  & -1.78    \\
30,000 & -2.46  & -1.11    \\
35,000 & -1.83  & -0.43    \\
45,000 & -0.57  & 0.92    \\
65,000 & 1.95 & 3.62 \\
 \hline 
\end{tabular}
\label{n_He}
\end{table}

The time spent on computational work increases and the convergence is more difficult when the complexity of the heavy elements is included in the \texttt{TLUSTY} atmosphere code \citep{hubeny1988, tlusty2011}. As a reference, each one of the 96 structures were computed and analyzed in a typical time scale of a day.

The grid of subdwarf-structure models is composed of eight effective temperatures ($T\textsubscript{eff}$: 10,000, 15,000, 20,000, 25,000, 30,000, 35,000, 45,000, and 65,000 K), each one computed in three surface gravities ($\log{g}$ [cgs]: 4.5, 5.5, and 6.5) and four different chemical abundances: solar and He-rich, solar and He-poor, halo and He-rich, halo and He-poor. 

The broadening of lines was treated carefully considering the Lyman and Balmer line profile tables from \cite{Tremblay2009}.  The high-frequency limit for lines that are taken into account in the opacity sampling was set to $6 \times 10^{11} \times T\textsubscript{eff}$ or the highest bound-free edge frequency, if greater. Opacity-sampling steps smaller than the thermal broadening were set. The microturbulent velocity was set to 10 km s$^{-1}$. The opacity-sampling approach was used to compute the superline cross sections with better accuracy, also, the iron peak lines were treated in a line-blanketed model.

The only exception to the description above was the cooler atmosphere models with 10,000 K which require a convective treatment \citep{Fontaine1981,Bergeron1992}; a mixing-length theory parameter equal to the pressure-scale height ($\alpha$\textsubscript{MLT} = 1.0) was used. For the 10,000 K case, the final structure model was computed in LTE without a microturbulent velocity. 

Most models were computed using the hybrid complete-linearization and accelerated lambda-iteration (CL/ALI) method \citep{CLALI1995, CLALI2003}, which is the default procedure for computing fully consistent, NLTE metal-line-blanketed atmosphere models in \texttt{TLUSTY} \citep{hubeny1988}. When the convergence was not achieved directly from CL/ALI we computed the atmosphere structure models using the Rybicki scheme \citep{Hubeny_and_Mihalas_book}, which is also suitable for treating high-temperature and high-gravity atmospheres, with better convergence behavior in some cases.

\begin{figure*}[ht!]
\includegraphics[width=\linewidth]{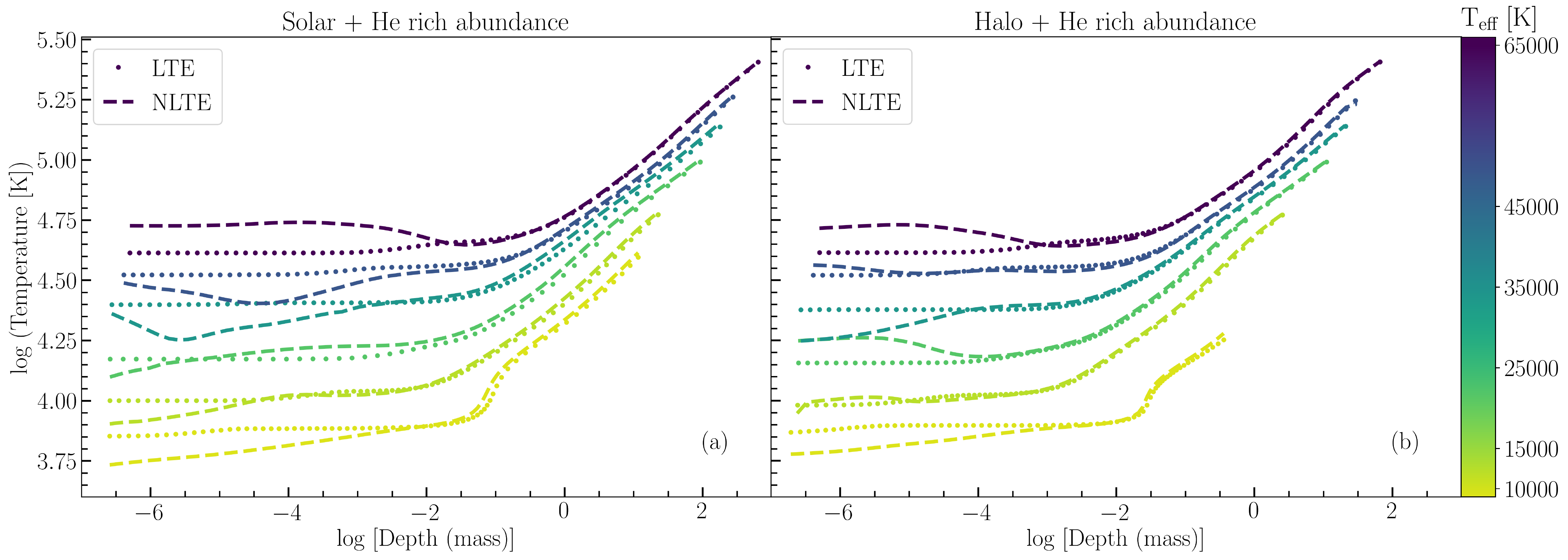}
\caption{Structure models of the temperature [K] as function of mass depth [g cm$^{-2}$] in a logarithmic scale computed using LTE (dotted lines) and NLTE (dashed lines) assumptions. The effective temperatures are shown in the colored scale from bottom (10,000 K) to top (65,000 K). a) These models have solar and He-rich abundances and $\log{g}$ [cgs] = 4.5. b) These models have halo and He-rich abundances and $\log{g}$ [cgs] = 6.5.
\label{fig:struc}}
\end{figure*}

The effective temperature as function of  the mass depth is shown in figure \ref{fig:struc}. The outermost mass depth corresponds to the Rosseland optical depth $\tau=10^{-7}$ while the innermost depth has $\tau=100$, logarithmically sampled in 70 layers. The color-bar scale indicates the effective temperature from bottom 10,000 K in yellow to top 65,000 K in purple. The hottest structures are more extended toward the inner thick region compared to the coolest model. The structure models in figure \ref{fig:struc} (a) have a surface gravity ($\log{g}$) equal to 4.5, so they are more extended if we compare them to the structures in figure \ref{fig:struc} (b), which represent more compact objects with $\log{g}$ equal to 6.5. The dotted lines represent the starting model computed using LTE assumptions, where we can see the almost isothermal behavior in the external region for models with different effective temperatures. On the other hand, the dashed lines represent the structure computed using the NLTE assumptions and line blanketing of the explicit species. In the latter, a small temperature inversion on their outer regions can be seen in some cases, which is important to the line-core profile formation. The inner region converges to the solution with minor differences if we compare the LTE and NLTE structures. Note that mass depths are shown, if we look at a specific line optical depth the differences between LTE and NLTE structure models are much more significant.

The structure models are highly sensitive to the chemical abundances via opacities, electron density, and mass density. The explicit atoms and ions included to compose the level and superlevels are the most important ingredients to compute detailed NLTE structure models. They have an impact on the cross section computation, justifying the careful choice of the most important ions for each model. 


\begin{figure}[ht!]
\includegraphics[width=\linewidth]{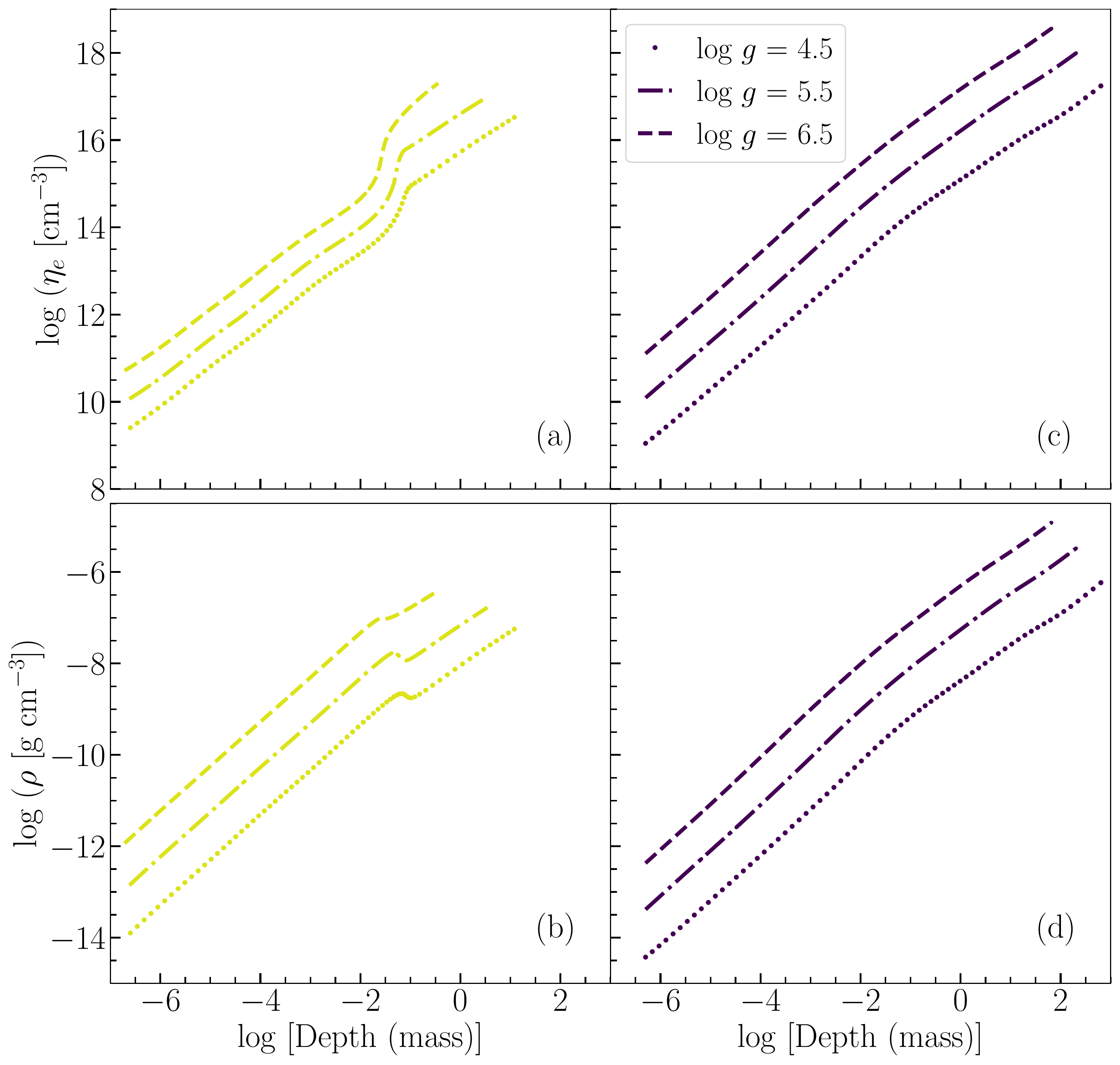}
\caption{Structure models of the electron density (top) and mass density (bottom) as function of mass depth [g cm$^{-2}$] in a logarithmic scale. We are comparing different surface gravity models computed using NLTE assumptions. These models have halo and He-rich abundances. a), b) $T\textsubscript{eff}$ = 10,000 K; c), d) $T\textsubscript{eff}$ = 65,000 K. 
\label{struc:dens}}
\end{figure}

The electron density and the mass density are shown in figure \ref{struc:dens}, having a similar profile as a function of the mass depth. The effects of the surface gravity in the atmosphere height are evident as already mentioned above. The coolest structure models such as the example of 10,000 K in figure \ref{struc:dens} a) and b) have an inversion on the density profiles near log [Depth (mass)] = -1, this is due to the effect of convection that was considered for these specific temperatures. The hot structure models as the example of 65,000 K in figure \ref{struc:dens} c) and d) have densities with a linear dependency of the mass depth.

\begin{figure}[ht!]
\centering
\includegraphics[width=\linewidth]{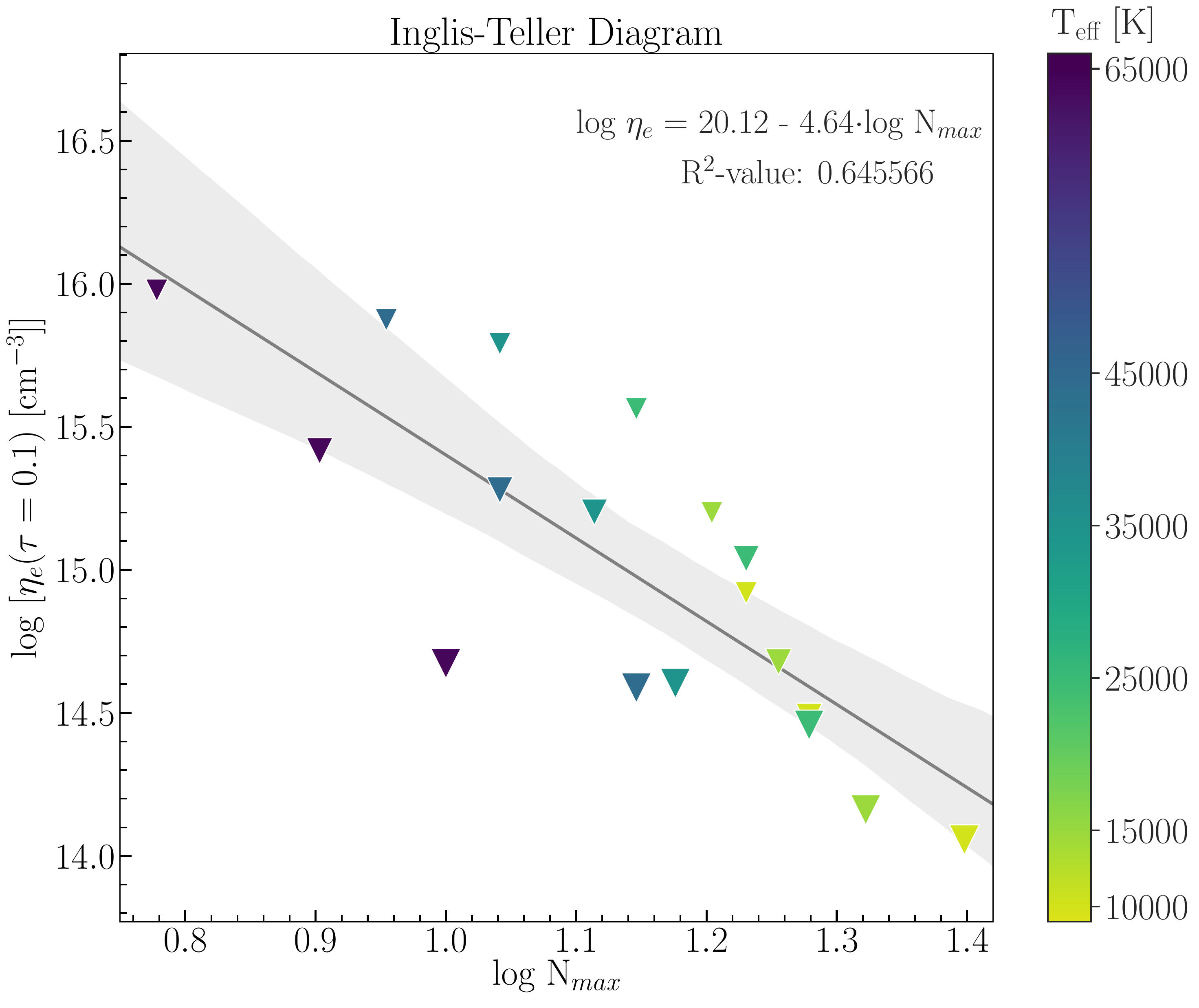}
\caption{An Inglis-Teller diagram of models with halo and He-rich abundances. \label{inglisteller}}
\end{figure}

An Inglis-Teller diagram for the grid was produced, which is a classical tool for evaluating model sequences' behavior over a gravity range. It involves the electron density and the maximum $n$ level that originates a distinguishable Balmer absorption line, which is useful for diagnosing our model's quality. We interpolated the model electron density to a specific optical depth $\tau$ = 0.1, counting the highest visible term in the final synthetic spectra. A linear least-squares fit of the electron density, and the maximum number of the absorption lines near the Balmer's discontinuity, is shown in figure \ref{inglisteller}.

We also performed convergence analyses for all the 96 subdwarf-structure models computed. The most critical convergence criterion is the magnitude of the relative changes of the components of the state vector, which is defined as a set of all structural parameters (e.g. temperature, particle number densities, and the mean radiation intensities in discretized frequency points) in a given discretized depth point \citep{tlusty2011}. The necessary condition for convergence is that the maximum relative change of all state vector components in all the 70 depths is smaller than $10^{-3}$. However, a supplementary condition is the conservation of the total flux concerning the total theoretical flux, $\sigma$T$\textsubscript{eff}^{4}$. 
The output parameters such as the number of depths, column mass, temperature, and densities in each depth and relative changes between iterations are used in the convergence analysis, as well as in the emergent flux in all frequency points used by the \texttt{SYNSPEC} code \citep{synspec2011} to build synthetic spectra. 

\section{Synthetic spectra} \label{sec:spec}

We computed the grid of synthetic spectra with the \texttt{SYNSPEC} code designed to synthesize the emergent spectra from atmosphere model structures. We used the NLTE atmosphere structure from \texttt{TLUSTY} models \citep{hubeny1988, tlusty2011} described in section \ref{sec:atm}. We considered NLTE assumptions for the evaluation of the level and superlevel populations. The line profiles were carefully treated in a special computation for H and He \citep{Lanz2003, Lanz2007}. The reference atomic line list is \cite{Coelho2014}, based on \cite{coelho2005} and line lists from R. Kurucz \footnote{\url{http://kurucz.harvard.edu/linelists.html}} and F. Castelli \footnote{\url{http://wwwuser.oats.inaf.it/castelli/linelists.html}} (see, e. g. \citealt{kurucz2017,castelli2004}. 

The hydrogen and hydrogenic lines are treated as a part of the continuum and their profiles are computed using tables from \citet{Tremblay2009}. The quasi-molecular satellites of L$\alpha$, L$\beta$, and H$\alpha$ ($\lambda$ = 1215.67,  1025.18 and 6562.79 \texttt{\AA}, respectively), are considered. In that case, additional input files containing the corresponding data were used \citep{allard2009}. 

\begin{figure*}[ht!]
\centering
\includegraphics[width=\linewidth]{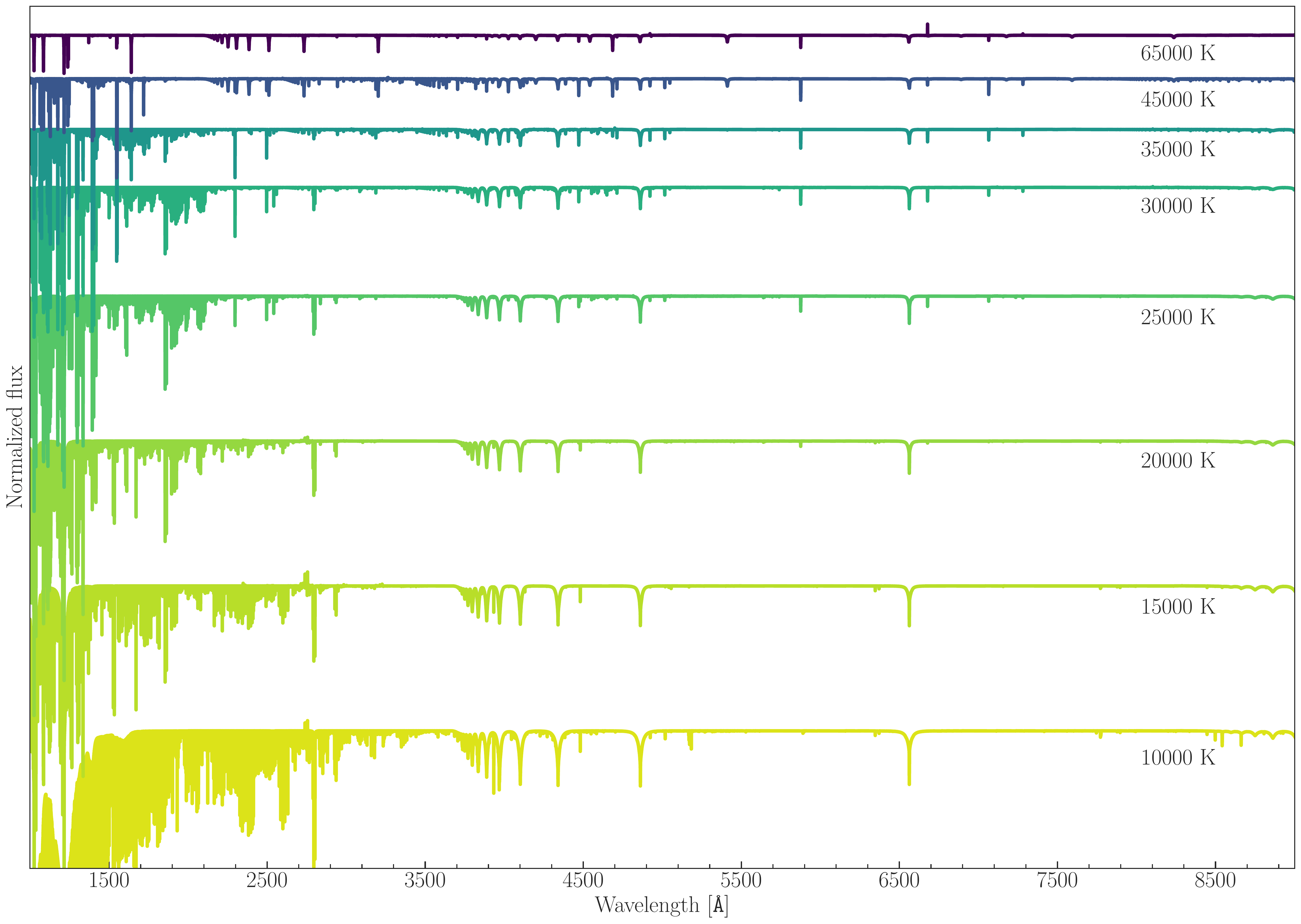}
\caption{Sample spectra with coverage from 1000 - 9000 \texttt{\AA} without instrumental broadening for different temperatures from top (65,000K) to bottom (10,000K), $\log{g}$ [cgs] = 5.5 with halo and He-rich abundances.\label{coverage}}
\end{figure*}

The four He \textsc{i} triplet lines, ($\lambda$ = 4026, 4387, 4471, and 4921 \texttt{\AA}) were treated using special line-broadening tables \citep{Barnard1974,shamey1969}. The He \textsc{ii} lines are treated as the approximate hydrogenic ion by analytical values of the Stark + Doppler profile \citep{nlte1994}, which improves the accuracy of the line profile for $T\textsubscript{eff}$ $>$ 10,000 K, and the line profiles are given by the Stark-broadening tables of \cite{schoning1989}. We are also considering Stark broadening computed by \cite{Tremblay2009}.

The spectral grid coverage is between 1000 -  10,000 \texttt{\AA} with steps of 0.01  \texttt{\AA}. The resulting spectrum was subsequently processed with the \texttt{ROTIN} code \citep{synspec2011}, which resamples the original synthetic spectrum but considering that no rotational velocity and no instrumental degradation have been taken into account.

\begin{figure}[ht!]
\includegraphics[width=\linewidth]{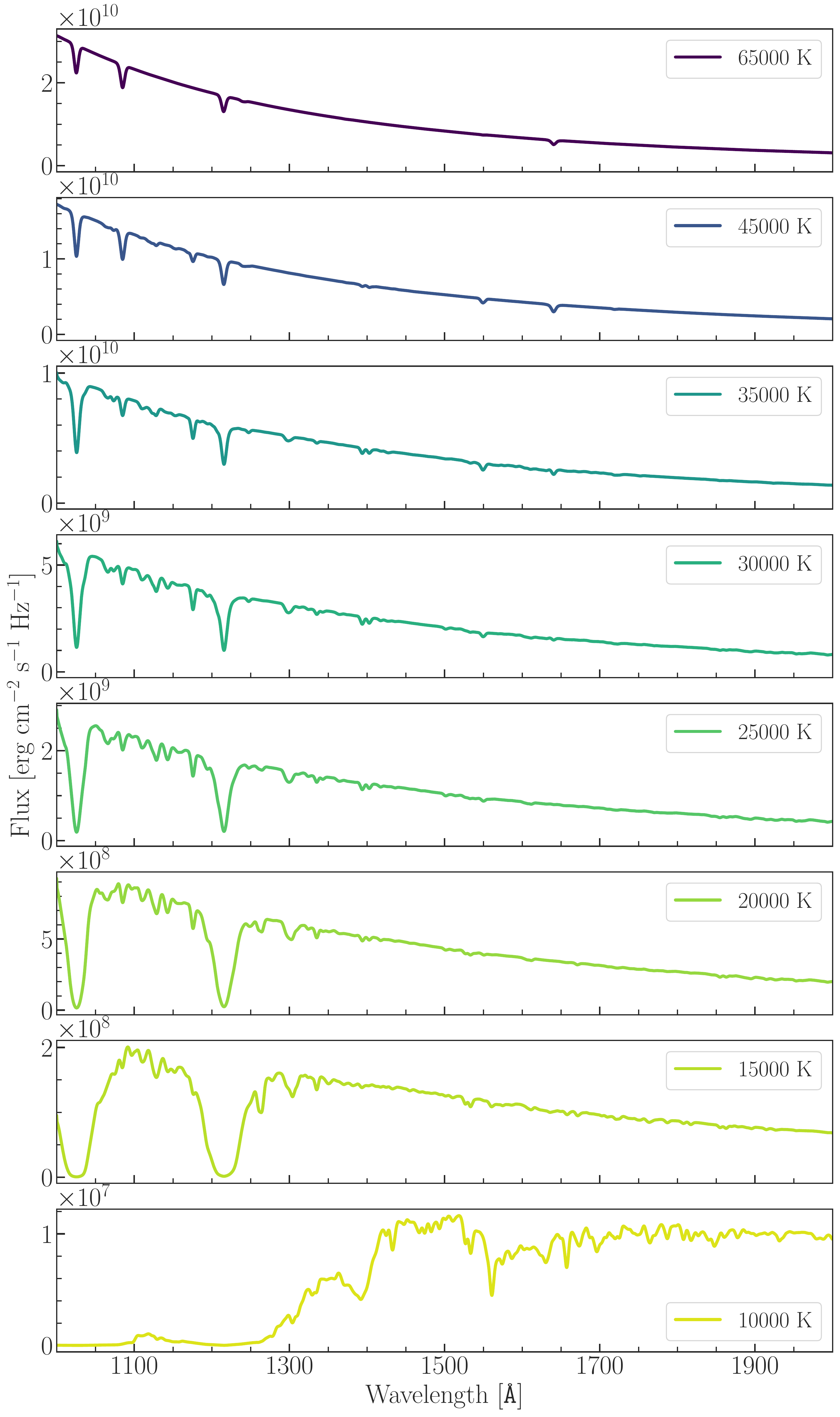}
\caption{Sample spectra in the UV region (1000 - 2000 \texttt{\AA}) with FWHM resolution = 5 \texttt{\AA} for different temperatures: from top (65,000K) to bottom (10,000K), and $\log{g}$ [cgs] = 5.5 with halo and He-rich abundances. \label{uvspec}}
\end{figure}

\begin{figure}[ht!]
\includegraphics[width=\linewidth]{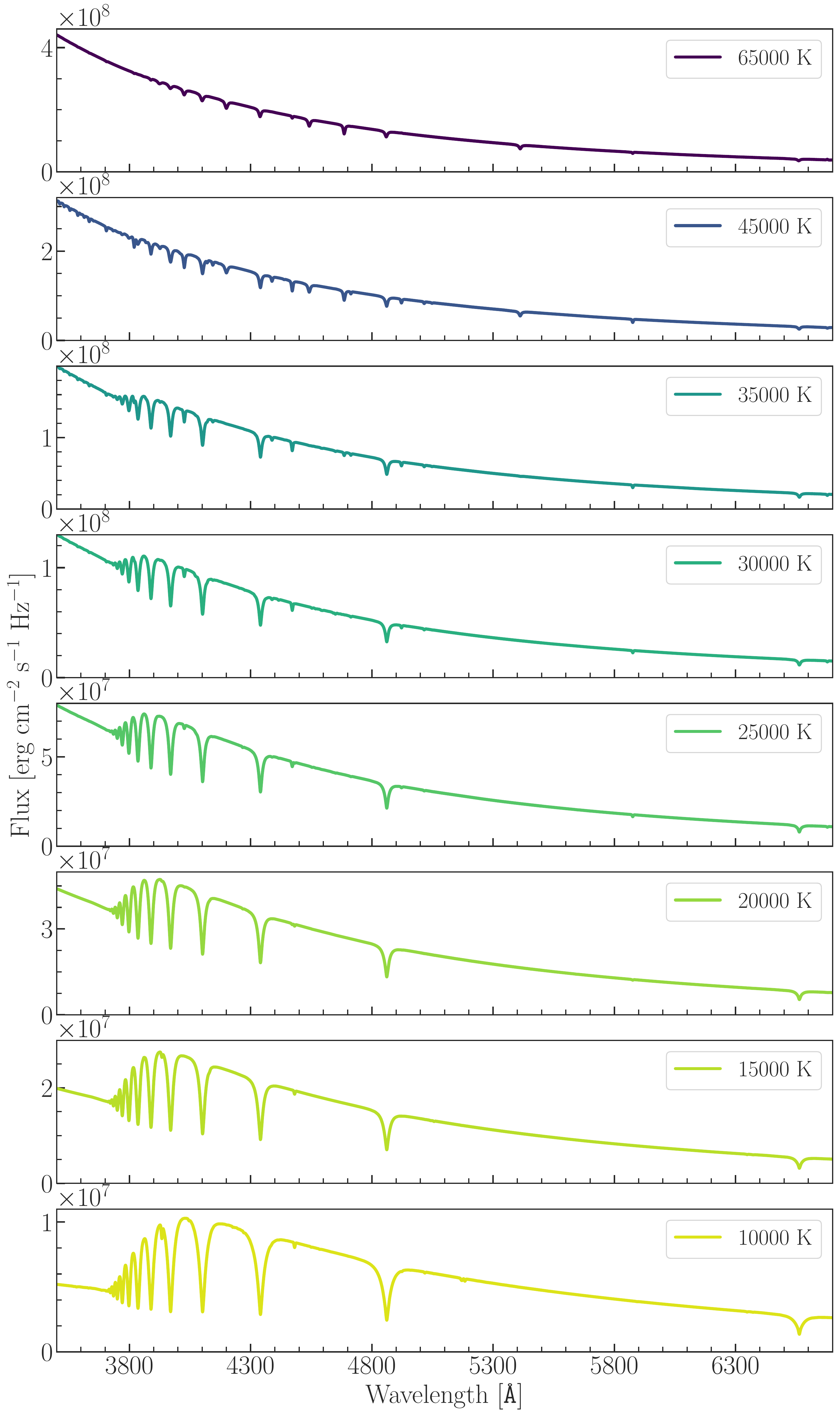}
\caption{Sample spectra in the optical region (3500 - 6750 \texttt{\AA}) with FWHM resolution = 5  \texttt{\AA} for different temperatures: from top (65,000K) to bottom (10,000K), and $\log{g}$ [cgs] = 5.5 with halo and He-rich abundances. \label{optico}}
\end{figure}

We show the normalized emergent flux (in arbitrary units) given as a function of wavelength for the spectral coverage from 1000 to 9000 \texttt{\AA} in figure \ref{coverage}. A forest of lines is seen at this particular sampling, where no instrumental degradation is included. They are synthetic spectra with $T\textsubscript{eff}$ equal to 65,000, 45,000, 35,000,  30,000,  25,000,  20,000, 15,000 and 10,000 K (from top to bottom) and $\log{g}$ [cgs] = 5.5 with halo and He-rich abundances. 

In figure \ref{uvspec} we can compare the different spectral types in the near-UV region between 1000 and  2000 \texttt{\AA}, where for visual effect it is degraded to a gaussian instrumental profile with full width at half maximum (FWHM) equal to 5 \texttt{\AA}. The hotter spectra have a bluer continuum and the He lines are dominant. The Stark broadening is more evident on the coolest spectra, where a stronger Lyman series is present, besides its quasi-molecular absorption features. 

In figure \ref{optico} we can compare the different spectral types on the Balmer break region between 3500 - 6750 \texttt{\AA}, where for visual effect the resolution is degraded by a gaussian instrumental profile with FWHM equal to 5 \texttt{\AA}. As expected, the H is mostly ionized and the Balmer series is weaker in the hotter spectra.  The coolest models are not favorable to the formation of strong He lines. Moreover, the spectra presented in figure \ref{optico} follow the He-abundance sequence from \cite{nemeth2012}, which shows low He abundance even for the He-rich sequence of cool subdwarfs.

\begin{figure}[ht!]
\includegraphics[width=\linewidth]{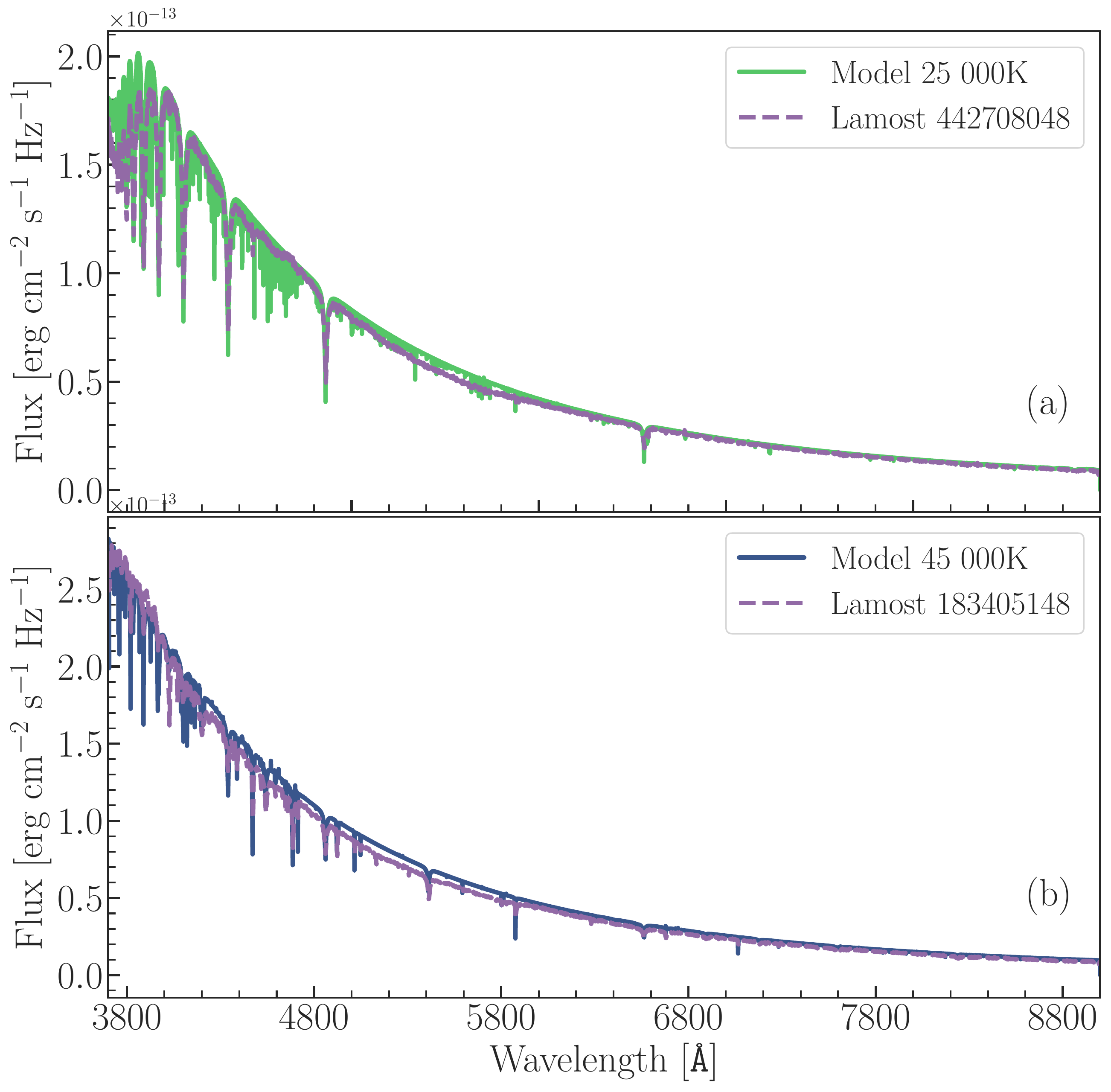}
\caption{Top: comparison between the observed spectrum of subdwarf Lamost 4427080 (dashed purple) and the model spectrum with $T\textsubscript{eff}$ = 25,000 K and $\log{g}$ = 5.5 (solid green). Bottom: comparison between the observed spectrum of Lamost 183405148 (dashed purple) and the model spectrum with $T\textsubscript{eff}$ = 45,000 K and $\log{g}$ = 5.5 (solid blue).}
\label{spec_opt}
\end{figure}

With the purpose of illustrating the grid, we present a simple comparison of model spectra with the spectra of the Lamost 442708048, Lamost 183405148, and HD 4539 subdwarfs. No interpolation in the grid was performed as the intent is not to determine exact abundances or stellar parameters, but rather to exhibit the grid potentialities. The targets were selected from the subdwarf catalog in \cite{Lei2019} and Hubble Space Telescope’s (HST)/Space Telescope Imaging Spectrograph (STIS) Legacy  Archive  data. They are located in the solar neighborhood. 
 
According to \cite{Lei2019}, the target Lamost 442708048 has $T\textsubscript{eff}$ = 26,620$\pm$70 K, $\log{g}$ = 5.53$\pm$0.01 [cgs], log ($n_{He}/n_{H}$) = -2.78$\pm$0.05, $E(B-V) = 0.018$ and, from the \cite{gaiaDR2} parallax, its distance is 317$\pm$5 pc. We performed a comparison with models near to these values, within the boundaries of our grid, as $T\textsubscript{eff}$ = 25,000 K, $\log{g}$ = 5.5 [cgs], [Fe/H] = 0, and log ($n_{He}/n_{H}$) = -3.09 (see figure \ref{spec_opt} (a)). 

Lamost 183405148 has, according to \cite{Lei2019}, $T\textsubscript{eff}$ = 46,270$\pm$330 K, $\log{g}$ = 5.88$\pm$0.04, log ($n_{He}/n_{H}$) = 0.29$\pm$0.01, $E(B-V) = 0.021$,  and from its parallax \citep{gaiaDR2} the distance is 333$\pm$9 pc. The 
closest model spectrum adopted corresponds to $T\textsubscript{eff}$ = 45,000 K, $\log{g}$ = 5.5 [cgs], [Fe/H] = 0, and log ($n_{He}/n_{H}$) = 0.92 (see figure \ref{spec_opt} (b)).

\begin{figure}[ht!]
\includegraphics[width=\linewidth]{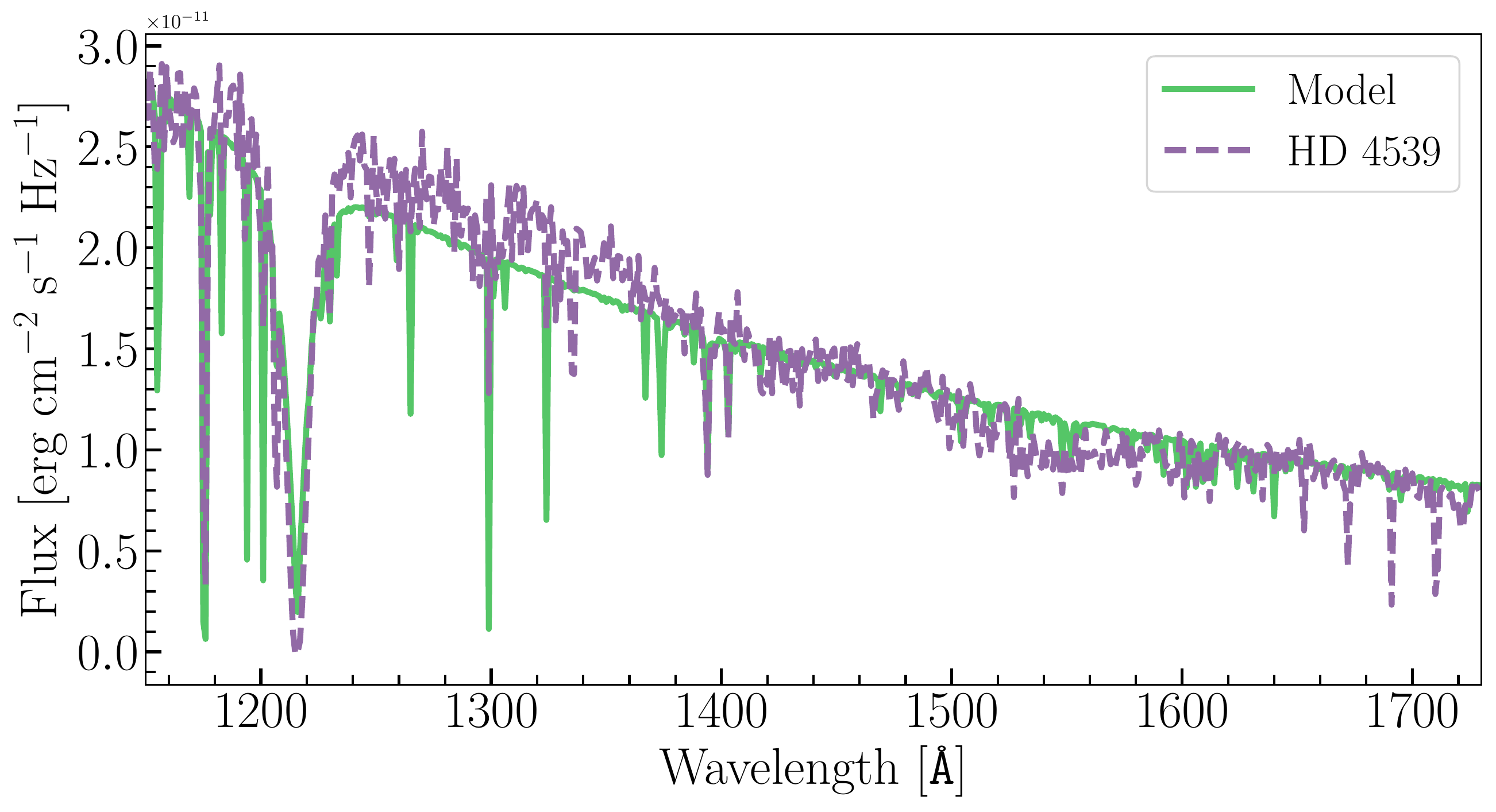}
\caption{Comparison between the observed HST/STIS UV spectrum of HD 4539 (dashed purple) and an interpolated model spectrum with $T\textsubscript{eff}$ 26,000 K and $\log{g}$ = 5.2 (solid green).} \label{spec_uv}
\end{figure}

HST/STIS Legacy Archive data on  the bright subdwarf HD 4539 were used to illustrate the models in the FUV. Parameters for this sdB are as follows:  $T\textsubscript{eff}$ = 26,000$\pm$500 K, $\log{g}$ = 5.2$\pm$0.1, log ($n_{He}/n_{H}$) = -2.32$\pm$0.05,  $E(B-V) = 0.04\pm0.01$ \citep{HDdata} and $d = 171.6\pm2.1$ pc (Gaia EDR3; \citet{gaiaDR3}). The models were scaled to match the continuum flux in the middle of each spectral range. He-poor branch and halo low-Z abundances were assumed here while $T\textsubscript{eff}$ and $\log{g}$ were linearly interpolated from model nodes to the literature values (see figure \ref{spec_uv}). Rotation, instrumental resolution, and exact chemical composition  were neglected, which would explain the significant differences found in the lines and continua offsets.
By using the Gaia distances and reddening values above, stellar radii compatible with typical values for subdwarfs in eclipsing binaries (e.g. \citet{sdradius} could be found. A line and/or continuum fitting of spectra can be performed with the XTGRID facility \citep{xtgrid}.

\section{Synthetic magnitudes} \label{sec:mag}

Synthetic magnitudes have been computed for several standard photometric bands to trend the grid's behavior in the color indices space and provide a comparison with photometry data. We used the filter response functions available on the Filter Profile Service at the Spanish Virtual Observatory  \footnote{\url{http://svo.cab.inta-csic.es/main/index.php}} to convolve our synthetic integrated fluxes in the AB and Vega systems. The filters were interpolated in the spectra wavelength steps within the Shannon-Whittaker scheme. Photon-counting integrated fluxes were assumed \citep{bessell1998}. Vega's zero-points were previously evaluated from Calspec's standard spectrum \citep{calspec2014}.

\begin{figure}[ht!]
\centering
\includegraphics[width=\linewidth]{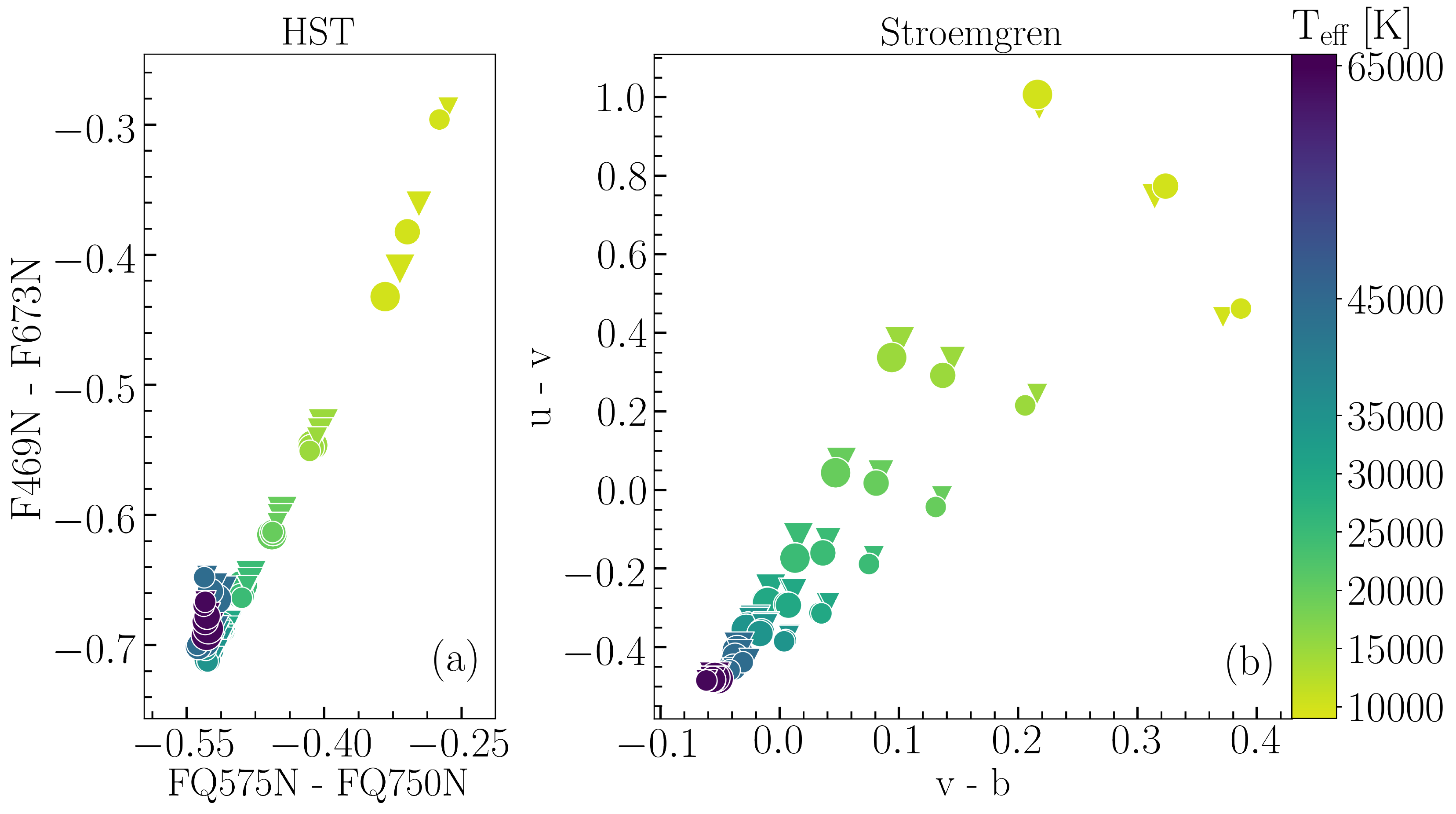}
\caption{Color-color diagrams with the scale showing the effective temperatures, while the point-size scale shows the surface gravities (larger to lower gravity (4.5) and smaller to the higher gravity (6.5)); the circles indicate solar abundances and down-pointing triangles the low halo metallicity. (a) Color-color diagram for HST/WFC3 indices F469N - F673N and FQ757N - FQ750N. (b) Color-color diagram for classical Strömgren \textit{u} (349.6nm) - \textit{v} (410.3nm) and \textit{v - b} (466.6nm) indices, aiming to probe the Balmer discontinuity.  \label{cmd}}
\end{figure}

Figure \ref{cmd} illustrate our models in two color-color panels: panel (a) was chosen to trace the overall continuum inclination, and panel (b) traces the behavior of the Balmer jump. In panel (a) we show F469N - F673N versus FQ757N - FQ750N colors of the HST/Wide Field Camera 3 (WFC3) photometric system.
The dependence on effective temperature is clear. Those indices measure how lower-temperature synthetic models have a flatter continuum in the optical region. The point-size scale represents the surface gravity, where larger symbols relate to lower gravity ($\log g$ = 4.5), and smaller symbols stand for higher gravity ($\log g$ = 6.5), which also reveal a linear dependence in the color space. Finally, the circles represent solar abundance and down-pointing triangles represent low halo metallicity, which has no clear separation on the diagram.

In figure \ref{cmd} (b) the Strömgren photometric system indices were used. Zero-points were derived from Vega $ubvy$ magnitudes \citep{hauck_merm1998} and its calibrated spectrum from the STScI CALSPEC database \citep{calspec2014}. The color-color diagram of figure \ref{cmd} (b) was constructed with \textit{u} (349.6nm) - \textit{v} (410.3nm) and \textit{v - b} (466.6nm) colors. It shows a temperature dependence of the Balmer break, which is more evident and also gravity-dependent for cooler atmospheres.

\begin{figure}[ht!]
\centering
\includegraphics[width=\linewidth]{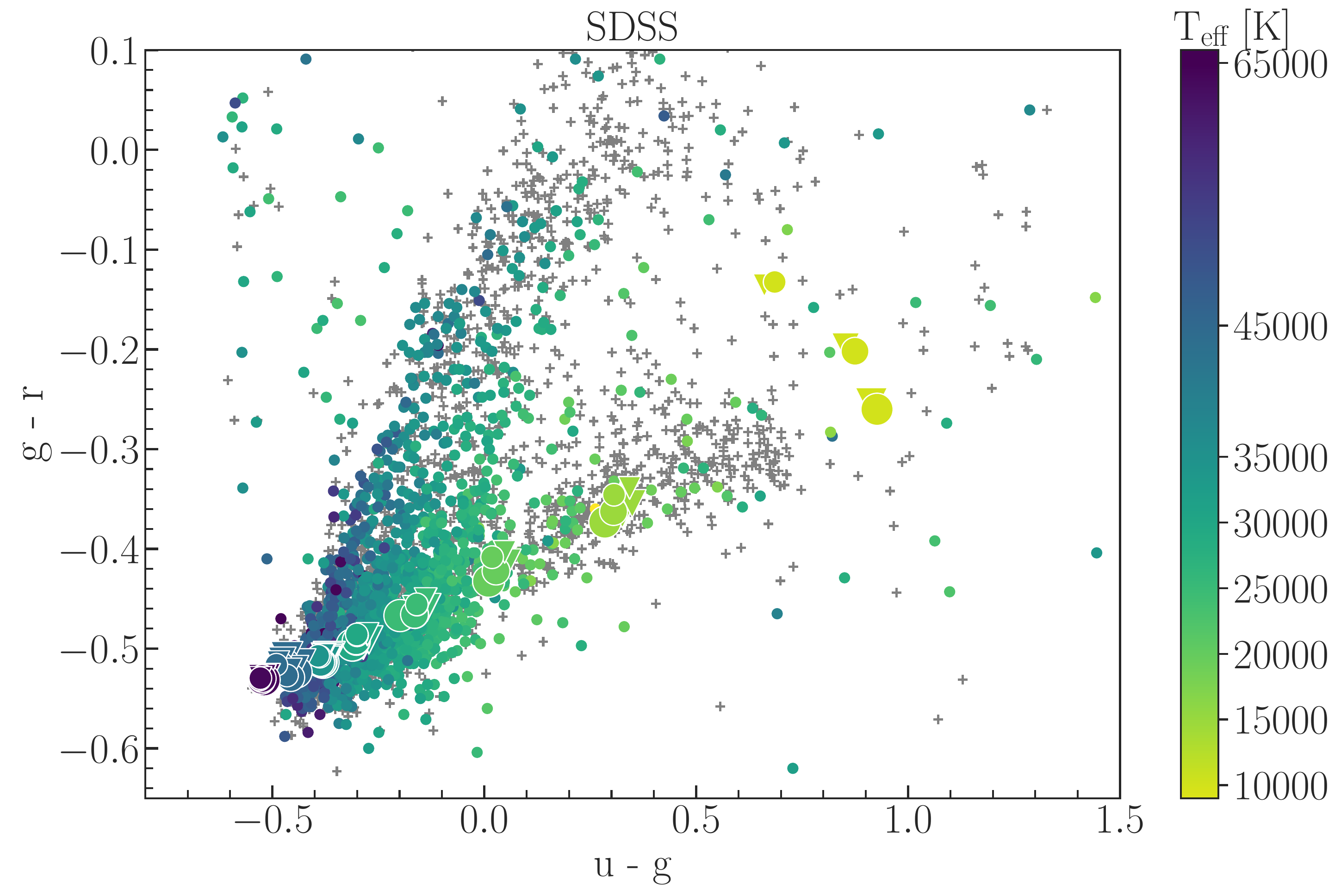}
\caption{Color-color diagram (\textit{g - r} vs. \textit{u - g}) composed of observational data on subdwarfs from SDSS (see text) with a hue scale representing determined effective temperatures, while the gray crosses shown an undetermined effective temperature. The synthetic colors from our grid are plotted with the same hue, style, and size scales as in figure \ref{cmd}.  \label{sdss}}
\end{figure}

The study of the 5874 hot subdwarf stars with Gaia Data Release 2 by \citealt{Geier2020} is the most complete sample of subdwarfs. From this catalog we selected data from 3450 targets (1482 with determined effective temperature and colors) observed by the Sloan Digital Sky Survey (SDSS) photometric system to compare with the grid synthetic magnitudes. The color-color diagram in figure \ref{sdss} is composed of \textit{g - r} and \textit{u - g} colors without any color cutoff. Observational data with a determined effective temperature are shown as dots scaled by hue, while those with an undetermined effective temperature are gray crosses. Our synthetic colors are displayed in the same hue, style, and size scales as in figure \ref{cmd}, matching the sdO, sdB, and sdOB previously classified by \citealt{Geier2020}. The upper data sequence represents the subdwarf composite binaries with main-sequence stars companions.

\section{Summary}\label{sec:conc}
We presented a grid of NLTE, fully blanketed theoretical spectra and synthetic photometry for hot and moderately cool subdwarf stars. The atmosphere models were computed considering line blanketing of H, He, C, N, O, Ne, Mg, Al, Si, S, and Fe. The effective temperatures are $T\textsubscript{eff}$ = 10,000, 15,000,  20,000, 25,000, 30,000, 35,000, 45,000, and 65,000 K, while the surface gravities are $\log{g}$ [cgs] = 4.5, 5.5, and 6.5. The two representative chemical abundances are solar and Galactic halo, each one with two extreme scenarios for He-rich and He-poor stellar atmospheres. 

The main differences between LTE and NLTE atmosphere structures are shown for these objects. They have significant differences in the outermost atmospheres, leading to distinct line-core profile formation. 

The complete high-resolution spectral synthesis is performed from the UV to near-IR (1000 to 10,000 \texttt{\AA}) in 0.01 \texttt{\AA}
steps. We provided an illustrative analysis for the UV and the optical regions by comparing our models with observed spectra from LAMOST and HST/STIS  Legacy  Archive data.

The behavior of the color indices were analyzed using the HST/WFC3 and the Strömgren photometric systems. A clear separation in effective temperature can be seen, as well as gravity for lower-temperature models, as provided by the Balmer discontinuity. We also matched our synthetic magnitudes against SDSS subdwarf data with fair agreement. 

These results pave the way for both spectroscopic and photometric analyses of fundamental parameters in isolated or binary objects which, in turn, may provide a more detailed insight into the atmosphere models themselves. In addition, classical stellar population synthesis analysis can make use of the homogeneous spectral grid to better understand the blue and UV properties of old stellar populations. The full spectral grid and synthetic indices are available in the IAG-USP $^2$, SVO $^3$, and Vizier $^4$ databases.

\begin{acknowledgements}

We thank Ivan Hubeny who kindly released the new versions of \texttt{TLUSTY} 208 and \texttt{SYNSPEC} 54 and boosted the discussions about the convective models. We thank the anonymous referee for refereeing this paper and for the suggestions that improved the work.

This study was financed in part by the Coordenação de Aperfeiçoamento de Pessoal de Nível Superior - Brasil (CAPES), Finance Code 001. M.P.D. acknowledges support from CNPq under grant $\#$305033. P.C. acknowledges support from Conselho Nacional de Desenvolvimento Cient\'ifico e Tecnol\'ogico (CNPq) under grant  $\#$310041/2018-0 and 
from Funda\c{c}\~{a}o de Amparo \`{a} Pesquisa do Estado de S\~{a}o Paulo (FAPESP) process $\#$2018/05392-8. 

This research has made use of the SVO Filter Profile Service (\url{http://svo2.cab.inta-csic.es/theory/fps/}) supported from the Spanish MINECO through grant AYA2017-84089.
This work has made use of data from the European Space Agency (ESA) mission
{\it Gaia} (\url{https://www.cosmos.esa.int/gaia}), processed by the {\it Gaia}
Data Processing and Analysis Consortium (DPAC;
\url{https://www.cosmos.esa.int/web/gaia/dpac/consortium}). Funding for the DPAC
has been provided by national institutions, in particular the institutions
participating in the {\it Gaia} Multilateral Agreement.

\end{acknowledgements}

\bibliography{library}{}
\bibliographystyle{aasjournal}



\end{document}